\documentclass{article}

\usepackage{PRIMEarxiv}

\usepackage[utf8]{inputenc} 
\usepackage[T1]{fontenc}    
\usepackage{url}            
\usepackage{booktabs}       
\usepackage{amsfonts}       
\usepackage{nicefrac}       
\usepackage{microtype}      
\usepackage{lipsum}
\usepackage{fancyhdr}       
\usepackage{graphicx}       
\graphicspath{{assets/}}     
\usepackage[utf8]{inputenc}
\usepackage[T1]{fontenc}

\usepackage{xargs}                      
\usepackage[pdftex,dvipsnames]{xcolor}  
\usepackage[colorinlistoftodos,prependcaption,textsize=tiny]{todonotes}
\usepackage[printonlyused, nolist]{acronym}
\usepackage{graphicx}
\usepackage{amsmath}
\usepackage{amsfonts}
\usepackage{bbold}
\usepackage{array}
\usepackage[numbers]{natbib} 
\usepackage{authblk}

\usepackage{tikz}
\usepackage{booktabs}
\usetikzlibrary{shapes.geometric, arrows}

\title{Automated Demand Forecasting in small to medium-sized enterprises
}

\author[1,*]{Thomas Gärtner} 
\author[1]{Christoph Lippert}
\author[1,2]{Stefan Konigorski}
\affil[1]{Digital Health Center, Hasso Plattner Institute for Digital Engineering, University of Potsdam, Germany}
\affil[2]{Hasso Plattner Institute for Digital Health at Mount Sinai, Icahn School of Medicine at Mount Sinai, New York, United States of America
}
\affil[*]{email: thomas.gaertner@hpi.de}

\begin{document}
\maketitle

\begin{abstract}
In response to the growing demand for  accurate demand forecasts to optimize production, purchasing, and logistics, this research proposes a generalized automated sales forecasting pipeline tailored for \acf{SME}.
Unlike large corporations that benefit from the expertise of data scientists for sales forecasting, \acp{SME} often lack such resources. 
In our research, we developed a comprehensive forecasting pipeline that automates time series sales forecasting, encompassing data preparation, model training, and selection based on validation results, thereby providing \acp{SME} with an advanced planning tool.

The development included two main parts: \textit{model preselection} and \textit{forecasting pipeline}. 
In the first phase, several state-of-the-art methods were tested on a showcase dataset to identify six models as suitable candidates for the pipeline. As final models, ARIMA, SARIMAX, Holt-Winters Exponential Smoothing, Regression Tree, Dilated Convolutional Neural Network, and Generalized Additive Model were selected. 
We also included an ensemble prediction of the models. 
Long-Short-Term Memory was not included in the pipeline as it did not achieve the desired prediction accuracy for the 18-month horizon, and Facebook Prophet, while robust, was excluded due to compatibility issues with the production environment.
In the second phase, the proposed forecasting pipeline was tested with \acp{SME} from food and electric industry, revealing variable model performance across different companies. 
One company, due to its project-based nature, saw no benefit, while the others experienced realistic and superior sales forecasts compared to naive estimators. 
Our findings suggested that no single model is universally superior; rather, an array of models, when integrated within an automated validation framework, can significantly enhance forecasting accuracy for \acp{SME}. 

The results emphasize  the importance of model diversity and automated validation to address unique needs of each business.
This research contributes to the field by giving access to state-of-the-art sales forecasting tools, enabling \acp{SME} to make data-driven decisions with improved efficiency.
\end{abstract}


\section{Introduction}

 
Within the last decades, companies were able to collect valuable data across the sales, production, logistics and procurement through the digitization of the supply chain.
\ac{ERP} systems enable a fully digital mapping of most of the processes within the company, utilized for documentation, and enable optimization by minimizing the resources.
One of the most crucial parts in companies focusing on production is having accurate sales forecasts to improve procurement, allocate resources, make data driven decisions, and estimate the expected revenue.
Overall, good sales predictions should reduce the overproduction and possible waste of resources by minimizing safety stock quantity as well as reducing the risk of being out of stock. 
While bigger organizations with high sales have resources to employ data analysts developing customized methods to get value out of the data, \acfp{SME} often do not have such (financial) resources.
With that, \ac{ERP} systems with integrated methods for data analysis and sales figures forecasting can be beneficial and make them more competitive as they allow to optimize internal processes and reduce costs.

Integrating methods for automated time series forecasting in \ac{ERP} systems for sales figures holds several challenges.
Various methods for time series forecasting were developed in the last decades with different strengths and limitations. \citep{vairagade_demand_2019, taylor_forecasting_2017, hassan_el_madany_procurement_2022} 
Results have shown that hat there does not exist a universally best model and that the best model is highly dependent on the products and the companies. 
As the methods can directly build on the data base from the \ac{ERP} system, the data preprocessing can be standardized quite easily.
Nevertheless, data maintenance to define the measurements and generating high-quality historical data differs between companies.
Companies with different specialisations have different requirements ranging from expiration dates within food industry to project-based production in especially construction industry. 
Generalizing those requirements leads to complex use cases and different uses of the forecasts.
While most companies have strong domain knowledge, they often lack experience with machine learning methods.
With that, any automated analysis of the data should be as simple as possible, but powerful at the same time.
Because \acp{SME} can produce a wide range of products, the methods need to be highly efficient in their computation time to deliver results in a reasonable time frame.
Each company have different times for production and procurement.
With that, the focuses of product planning could be on a weekly or monthly basis and could be up to 2 years to get a rough estimation of future trends.

To advance the field of demand forecasting for \acp{SME}, in this paper, we introduce a generalized automated demand forecasting pipeline that integrates automatic validation and model selection. This pipeline was designed to simplify and streamline the forecasting process, allowing \acp{SME} to leverage advanced predictive models without the need for specialized data science expertise. One of the key contributions of this research is the incorporation of a comprehensive model evaluation framework, which ensures that the most suitable forecasting model is automatically selected based on validation results. By focusing on monthly long-term predictions, the pipeline provides a robust, scalable solution that adapts to various business contexts, enhancing forecasting accuracy and operational efficiency for \acp{SME}.


The paper is structured as follows. 
First, we provide a general introduction followed by related work. Next, in Section \ref{sec:model-preselection}, we discuss an evaluation of different models on a showcase dataset in order to identify promising forecasting approaches. 
Afterwards, we introduce the data analysis pipeline in Section \ref{sec:forecasting-pipeline}. 
In Section \ref{sec:user-test}, we perform a user test with a pilot group consisting of 5 companies to evaluate the pipeline under real-world conditions. 
Finally, we discuss the results, the potential impact on the business and potential next steps to further improve the pipeline in Section \ref{discussion}.

\section{Related Work}

In the field of time series analysis for sales and demand forecasting, various methods have been developed and investigated. 
They can be clustered into two general groups of classical statistical models such as \ac{ARIMA} or \acl{ES}, and machine learning models including \aclp{RNN} or \acl{LSTM}.

Starting with classical statistical methods, \ac{ES} were introduced  for demand forecasting in 1956. \citep{brown_exponential_1956} 
Holt and Winters  further developed the model to the well-known \ac{HWES} incorporating seasonality. \citep{holt_forecasting_2004, winters_forecasting_1960}
In 1970, Box and Jenkins introduced a Method for estimation in \ac{ARIMA} models. \citep{box_time_1970}
In 1990, this method was reviewed and seasonality and trends as components in a structural model closer to the underlying economical concepts was introduced. \citep{harvey_forecasting_1990}
The use of tree-based methods such as \acf{RF}  or \ac{XGBoost} for time series forecasting was explored in different learning tasks, for instance in gold price prediction, blood sugar prediction in diabetes type-I patients as well as demand forecasting for supply chain management \citep{el_time_2021, syafrudin_personalized_2022, vairagade_demand_2019}.
In 2021, tree based methods such as \acf{RF}, \acf{XGBoost} and \ac{ARIMA} were compared in a study, to evaluate their performance on gold price prediction.
In their study \ac{RF} was able to outperform others, but they suggested a hybrid approach of \ac{ARIMA} and Deep Learning for forecasting. \citep{el_time_2021}

In 1997 \ac{LSTM} were introduced, a model trained through recurrent backpropagation capable of connecting information in long time lags \citep{hochreiter_long_1997}.
With similar properties, \acp{GRU} were introduced \citep{cho_properties_2014}.
\acp{CNN}, prominent in the field of computer vision, can be used for time series forecasting by using 1-dimensional kernels \citep{koprinska_convolutional_2018}. 
In 2018, a special architecture called dilated \acp{CNN} were used for time series forecasting.\citep{borovykh_dilated_2018}

A comparative study, using iterative predictions and direct predictions, found that direct Artificial Neural Networks outperformed other methods in their study.
However, direct forecasting raised new challenges and, for example, more data with longer history is required. 
\citep{hamzacebi_comparison_2009}

With Facebook Prophet, a data analytic pipeline including an analyst in the loop to improve the interpretation and debugging of the model was introduced. The model itself is a Bayesian \acf{GAM} including trend with change points, seasons, and features like holidays or events. \citep{taylor_forecasting_2017}

To generate demand forecasts for planning, Sugiarto et. al. used \ac{HWES} intergrated in an \ac{ERP} system. \cite{sugiarto_sales_2017}
In contrast, Hassan El Madany et.al. implemented a pipeline integrated in an \ac{ERP} system for procurement forecasting using \ac{ARIMA} and \ac{LSTM} determining the price for various products. \citep{hassan_el_madany_procurement_2022}
A case study about the usage of machine learning in companies highlighted various challenges using demand forecasting in an \ac{ERP} system, and comparing different approaches for predictions. \citep{grobler-debska_research_2021}

\section{Model Preselection}\label{sec:model-preselection}

Many different models are available for time series forecasting.
To identify the most suitable models for our analysis pipeline, we conducted a first analysis on a showcase data set to evaluate and compare the models.
Based on the performance and implementation details, we selected the most promising model for the next step.

\subsection{Show Case Data Set}

\begin{figure}[ht!]
    \centering \includegraphics[width=0.8\textwidth]{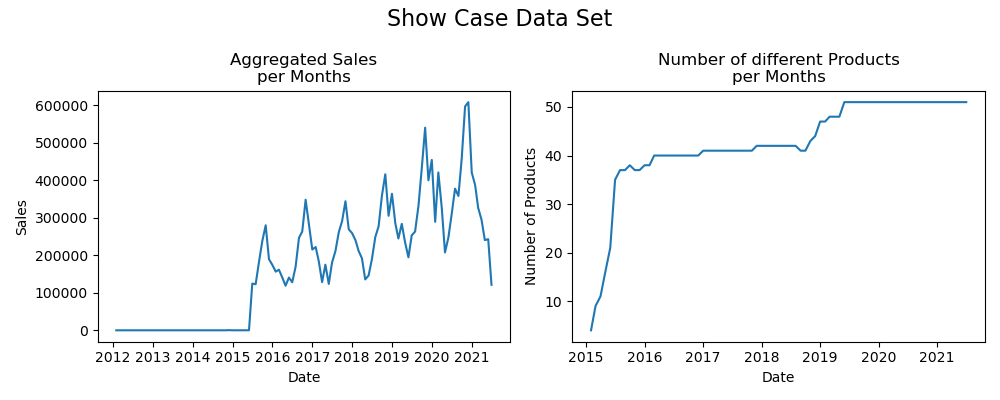}
    \caption{Overview over the showcase data set for model preselection. It shows the number of total sales for all 51 products (left) and the number of different products sold per month (right). Notably, new products were introduced in 2019 leading to an increase in sales at the same time.}
    \label{fig:OverviewShowCaseDataSet}
\end{figure}

The showcase data set contained 51 products provided by one company with a historical data of up to 6 years.
37 products ($72,5\%$) had data available at least from the beginning of 2016, while others were launched later in 2019. 
Figure \ref{fig:OverviewShowCaseDataSet} gives an overview of the aggregated sum of sales and the number of unique products on a monthly basis.

\begin{figure}[ht!]
    \centering
    \includegraphics[width=\textwidth]{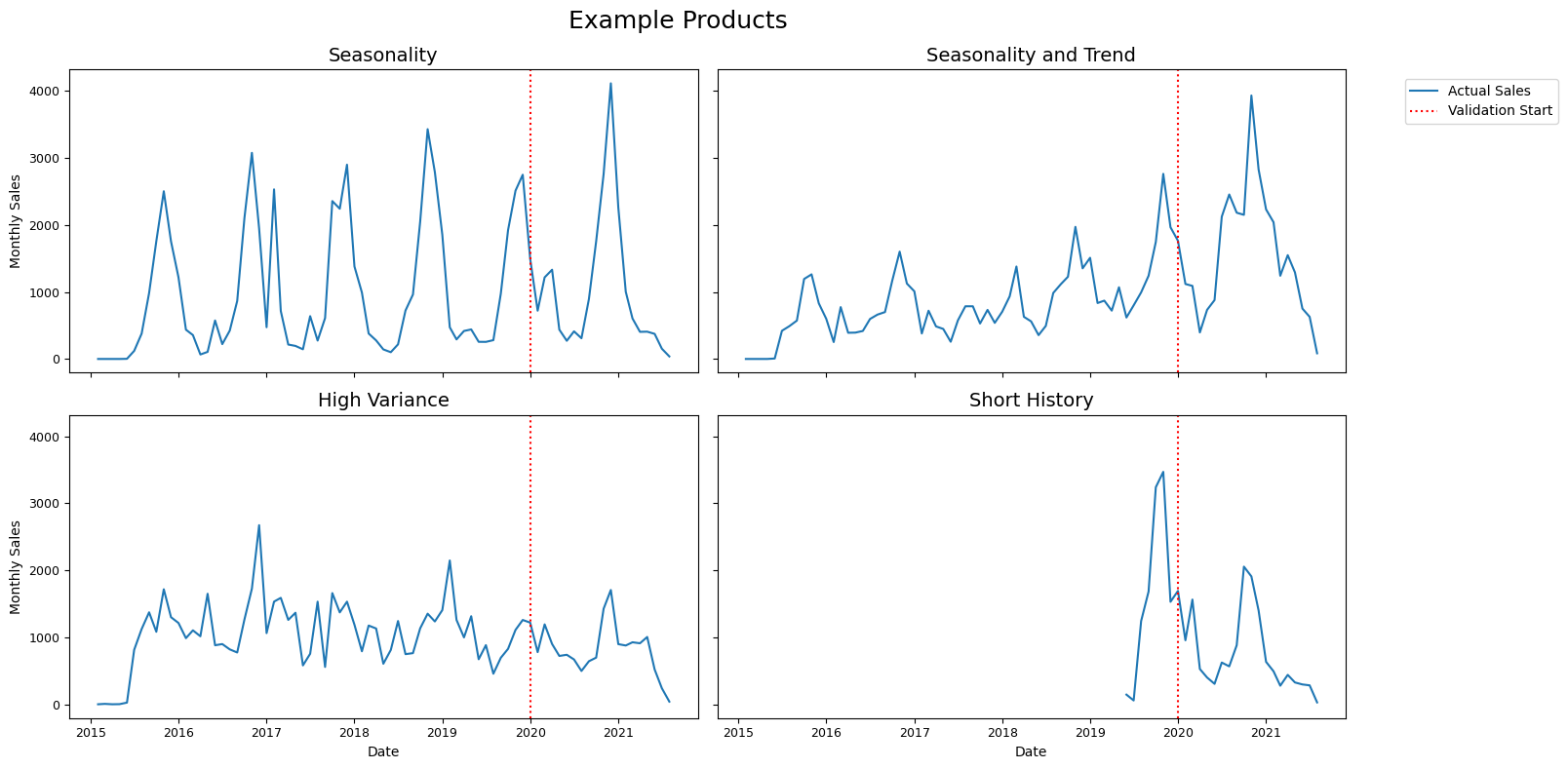}
    \caption{Sales for 4 different products of the showcase dataset.}
    \label{fig:SampleProducts}
\end{figure}

The products were selected by the company to showcase a variety of business use cases in the data set.
That includes products sold B2B as well as products sold B2C affecting the seasonality and the noise within the data.
Figure \ref{fig:SampleProducts} shows four different product histories as examples for different product behavior such as seasonality, trend, variance or observation time. 
For presenting the results, the data was rescaled, so that the average sales are equal to 1000 per month to de-identify the products. 

\subsection{Evaluated Methods}

Within the model preselection, we investigated 33 different models and model specifications ranging from traditional statistical methods to deep learning methods and compared their performances based on \acs{MAPE} and \acs{nRMSE}.

\subsubsection*{Autoregressive Models}

We fitted seasonal autoregressive integrated moving average (SARIMA) models with automatic selection of $(p,d,q)$ and $(P,D,Q)$-parameters, representing the order of autoregression, nonseasonal and seasonal differencing, and the size of nonseasonal and seasonal moving average windows, respectively. We utilized a grid search technique implemented in \texttt{pmdarima}, as detailed in \cite{smith_pmdarima_2017} with default parameters. This method systematically explores various combinations of parameters to identify the best fit for the time series data.

\subsubsection*{\acl{ES}}

Exponential smoothing is a time series forecasting method used for smoothing data points to identify trends. 
It applies a series of weights that decrease exponentially over time, making recent observations more influential in forecasting future values. 
Holt-Winters exponential smoothing extends this concept by addressing seasonality in addition to level and trend. 
It incorporates three smoothing parameters: level, trend, and seasonal components, which adapt the method for data with seasonal fluctuations. 
This enhancement allows \ac{HWES} to provide more accurate forecasts for data with underlying seasonal patterns compared to basic exponential smoothing.
In our experiments, we used the implementation from \texttt{statsmodels} with default parameters. \citep{hyndman_forecasting_2014}

\subsubsection*{\acl{RF}}

Tree-based methods are popular in time series forecasting due to their effectiveness in handling various types of data and their ability to capture complex patterns. 
Decision trees, \acf{RF}, and boosting methods such as \acf{XGBoost} are commonly used in both academia and industry for predictive tasks \citep{el_time_2021}.
In our study, the implementations from \texttt{sklearn} of \ac{RF} and \ac{XGBoost} as detailed in \cite{chen_xgboost_2016} were used.
For that, we defined the last year as input to predict the next time point.

\subsubsection*{Deep Learning Models}

Deep learning has become increasingly prominent in time series forecasting due to its ability to model complex and nonlinear relationships in data. 
\acfp{MLP}, \acfp{CNN}, \acfp{GRU}, and \acfp{LSTM} are particularly effective for this purpose. 
\acp{CNN} excel in identifying hierarchical patterns through their convolutional layers, making them suitable for time series that exhibit spatial dependencies.  \citep{koprinska_convolutional_2018} 
\acp{GRU} and \acp{LSTM}, both variants of recurrent neural networks, are adept at capturing temporal dynamics due to their memory cells, which help maintain information over longer sequences. \citep{hochreiter_long_1997}
These methods have been extensively applied across various sectors in both academia and industry, demonstrating strong performance in forecasting tasks involving sequential data.

In our model preselection, the deep learning methods were trained from scratch combined on all products with random initial weights for a maximum of 100 epochs, but early stopping was applied if the loss did not improve over five epochs. 
For all \acp{CNN}, we used the same architecture and learning parameters identified through hyperparameter tuning beforehand on the same data set. 
As the data set only containted 51 products, we also report the performance of a pretrained \ac{CNN} model on weather data. 
Furthermore, we compared different model specifications such as a dilated \ac{CNN} by \cite{borovykh_dilated_2018}. 
Another approach to reduce the risk of overfitting is fixing the weights of last layer to reduce the number of trainable parameters. This model was called \textit{\acp{CNN} (fixed)}. 
As we have daily data available, we also trained the models on granular data to predict the next day and aggregate the predictions to monthly forecasts in a second step.
Those models were called \textit{DAY} models.
Altogether, we trained and evaluated six different architectures of \acp{CNN} in the first model selection step namely \textit{\acs{CNN}}, \textit{\acs{CNN} (Dilated)}, \textit{\acs{CNN} (pretrained)}, \textit{\acs{CNN} (fixed)},
\textit{\acs{CNN} (Dilated)},\textit{\acs{CNN} (fixed, DAY)}, and \textit{\acs{CNN} (DAY)}.

Furthermore, we trained and evaluated three different architectures of \acp{LSTM}, one \ac{GRU} and one \ac{MLP} on the showcase dataset.

\subsubsection*{Facebook Prophet}

Facebook Prophet is a versatile forecasting tool designed for handling time series data that displays patterns on different time scales such as daily, weekly, or yearly seasonality. Developed by Facebook's data science team, Prophet is particularly user-friendly, making it accessible for analysts with varying levels of expertise. 
It excels in dealing with missing data and trend changes, and can incorporate holiday effects, which are often challenging for standard time series models. 
Their method combines configurable models with analyst-in-the-loop performance analysis.
In our analysis, we used their published python package with default parameters. 
\citep{taylor_forecasting_2017}

\subsubsection*{\acl{GAM}}

Following that, we developed a decompositional \acf{GAM} of the time series.
In this approach, we incorporate linear and exponential trends, seasonality, and user specific external features. 
Our model is an adaption of Facebook Prophet but instead of a Bayesian Model, we used a linear model for forecasting. 
Instead of a classical time series, we model the sales figures as a function of time, $f(t)$, with trend $T(t)$, seasonality $S(t)$, external features $X(t)$ at time point $t$ and polynomial spline functions with $\pi(t)$. With that, the model can be formalized as: 

$$
f(t) = \beta_0 + \beta_1 \cdot T(t) + \beta_2 \cdot S(t) + \beta_3 \cdot X(t) + \beta_4 \cdot \pi(t) + \epsilon,
$$

where $\epsilon$ is the error term following a normal distribution.

To determine trends, we used linear and exponential functions of time. 
For modeling  seasonality, we used Fourier transformation on monthly and weekly frequency.
The $\beta$ coefficients were estimated through coordinate descent with Lasso penalization. \citep{pedregosa_scikit-learn_2011} 

\subsubsection*{Ensemble}

The methods under consideration often exhibit tendencies toward overestimation or underestimation in their forecasts. 
To mitigate this and enhance robustness, we combined the predictions from an ensemble of statistical models, including \ac{HWES}, \ac{GAM}, \ac{ARIMA}, and \ac{XGBoost}. 
To integrate the forecasts from these diverse models, an appropriate aggregation function is required. 
In the first analysis, we employed \textit{mean} as well as \textit{median} functions to aggregate the predictions, aiming to balance the individual model biases and achieve more accurate overall forecasts.
We called the models Ensemble Mean and Ensemble Median respectivly.
We did not use a weighted mean as estimating robust weights in production can be complex.

\subsection{Experimental Set Up}

For analyzing the performance of the models, we splitted the data into train and test data. 
Data before January 1, 2020 was considered for training and data after January 1, 2020 was used for evaluation. 
That led to a short training history for 4 products with less than a year.
We trained all models described in the previous section. If not further specified, default parameters were used.
Next, we forecasted sales for the next 18 months starting from January 1, 2020, which increased the complexity of the forecasting task due to the long prediction horizon, which is essential in industry. 
As we had short historical data available, we used one-step ahead forecasting to use as much data as possible, and iterated over the prediction as the input for the next month.  

We assessed the model performances using \ac{nRMSE} as evaluation metric to compare the model performance across the products:

\begin{equation} \label{eq:RMSE}
\begin{split}
    \textit{RMSE}(y, \hat{y}) & = \sqrt{ \sum_{t} (y_t - \hat{y_t})^2}, \\
    \textit{nRMSE}(y, \hat{y}) & = \frac{\textit{RMSE}(y, \hat{y})}{y_{max} - y_{min}},
\end{split}
\end{equation}

where $y$ are the actual values and $\hat{y}$ are the predicted values for each time point $t$.

Additionally, we used the \acf{MAPE} defined as following for interpretation:

\begin{equation} \label{eq:MAPE}
MAPE = \frac{\sum_{t}^{n} \Big| \frac{y_t - \hat{y}_t}{y_t} \Big|}{n}, 
\end{equation}

where $n$ is the number of time points.

We analyzed the overall performance across all products by calculating the mean \ac{nRMSE} for each model, aggregating the results across all product time series. 

\subsection{Results}

Our findings revealed that the top-performing models predominantly belong to the deep learning category, including dilated \ac{CNN}, a \ac{CNN} without specifications, and a \ac{CNN} pre-trained on weather data as shown in Table \ref{tab:top10-models-preselection}.
A full list of model performances can be found in the Appendix Table \ref{sup-table:error-all}.
Tree-based boosting methods, \ac{GRU}, \ac{LSTM}, and \ac{MLP} all showed good performance, too, and were also ranked among the top 10 models, highlighting their robustness in time series forecasting.
Nevertheless, the average \ac{nRMSE} of the top-10 models are close to each other. 
The results of the \ac{MAPE} confirmed that the \ac{CNN} (Dilated) has the best model performance, deviating on average $33.21\%$ from the actual sales. Otherwise, the ranking of the models based on the \ac{MAPE} would lead to slightly different results though still quite similar. 

\begin{table}[!ht]
    \centering
    
        \begin{tabular}{lrr}
        \toprule
        Method           &  \acs{nRMSE} &  \acs{MAPE}         \\
        \midrule
CNN (Dilated)             &  0.3118 &  33.21 \% \\
CNN                       &  0.3119 &   37.0 \% \\
CNN (pretrained)          &  0.3137 &  35.05 \% \\
CNN (FIXED)               &  0.3245 &  36.81 \% \\
Tree-Based (XGBoost, log) &  0.3451 &  34.97 \% \\
Tree-Based (XGBoost)      &  0.3460 &  36.63 \% \\
GRU                       &  0.3502 &   47.3 \% \\
CNN (DAY)                 &  0.3567 &  36.95 \% \\
LSTM                      &  0.3571 &   45.0 \% \\
MLP                       &  0.3641 &  42.46 \% \\
        \bottomrule
        \end{tabular}
        
    \caption{Top-10 models sorted by the \acf{nRMSE} and the corresponding \acf{MAPE} averaged over all products.}
    \label{tab:top10-models-preselection}
\end{table}

\begin{figure}[ht!]
    \centering
    \includegraphics[width=\textwidth]{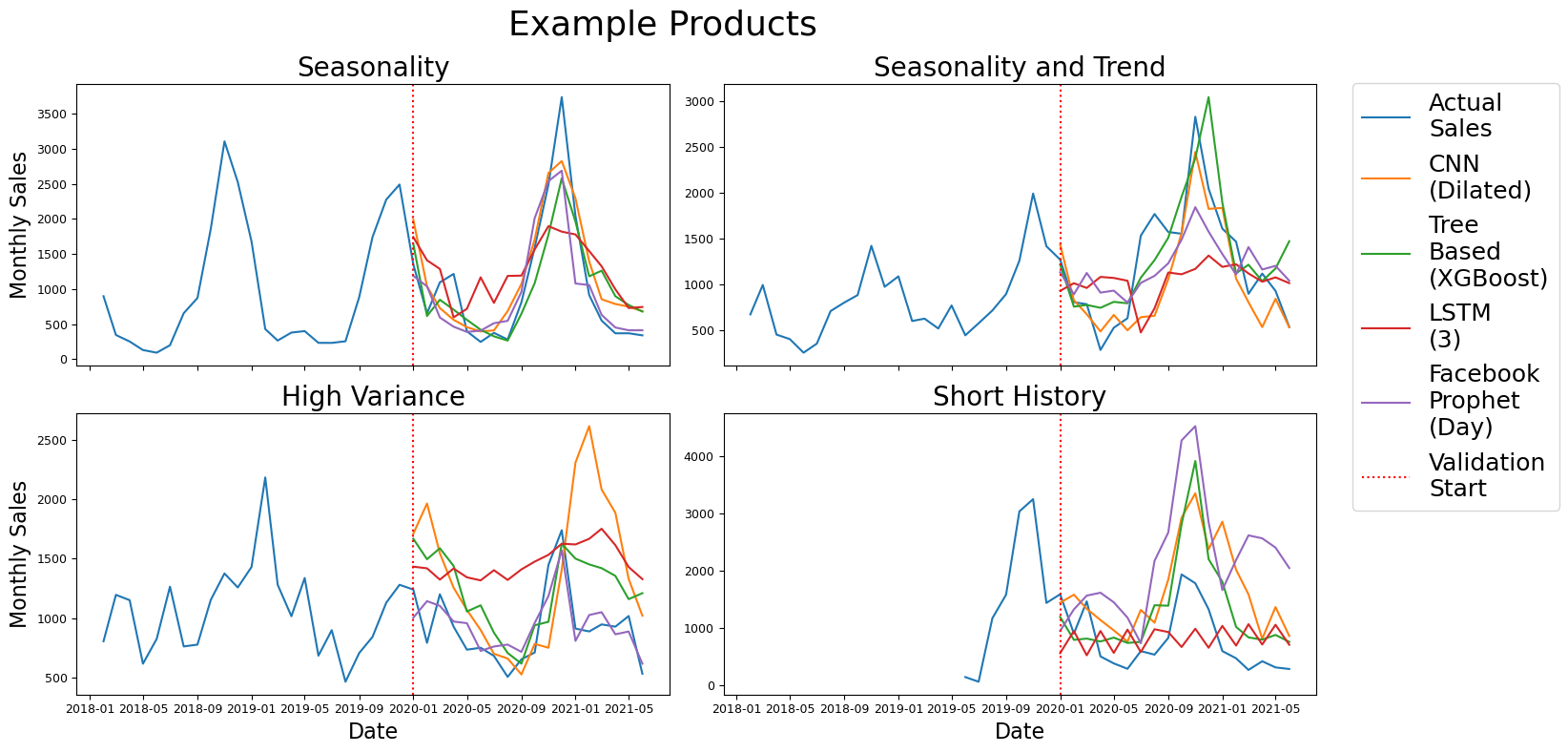}
    \caption{Example Sales figures for 4 different products with predictions by \textit{CNN (Dilated)}, \textit{Tree-Based (XGBoost)}, a version of \textit{\ac{LSTM}}, and \textit{Facebook-Prophet (Day)}.}
    \label{fig:SampleProductsWithPredictions}
\end{figure}

Figure \ref{fig:SampleProductsWithPredictions} illustrates four example products with different behaviour and the corresponding predictions by 4 different models such as \textit{CNN (Dilated)}, \textit{Tree-Based (XGBoost)}, \textit{\ac{LSTM}}, and \textit{Facebook-Prophet (Day)}. 
Inspecting the figure visually, it can be seen that all models, except for the \textit{\ac{LSTM}}, deliver reasonable predictions for data generated with seasonality, and generated with seasonality and trend, while \textit{Facebook-Prophet} tends to slightly underestimate the trend in this product. In this example product with high variance, \textit{\ac{XGBoost}} and \textit{Facebook-Prophet} were able to deliver reasonable predictions.
In this example product with a short history, all models beside the \textit{LSTM} expected a peak of sales at the end of 2020 larger then the actual sales.

\begin{table}[!ht]
    \centering
    \small{
        \begin{tabular}{lcccc}
        \toprule
        Product &  CNN (Dilated) &  Prophet &  LSTM &  XGBoost \\
        \midrule
        High Variance         &          0.377 &                   0.163 &     0.452 &                 0.304 \\
        Seasonality           &          0.124 &                   0.128 &     0.223 &                 0.133 \\
        Seasonality and Trend &          0.183 &                   0.189 &     0.303 &                 0.167 \\
        Short History         &          0.492 &                   0.879 &     0.412 &                 0.503 \\
        \bottomrule
        \end{tabular}
   
    }    
        \caption{\ac{nRMSE} for example products predicted with \textit{\acs{CNN}}, 
    \textit{Facebook Prophet (Day)}, 
    \textit{\acs{LSTM}} and 
    \textit{XGBoost}.}
    \label{tab:SampleProducts_MAPE}
\end{table}

The results of the visual inspection were confirmed by the \ac{nRMSE} shown in Table \ref{tab:SampleProducts_MAPE}. 
Additionally, the table showed that the prediction of products with a short history led to higher errors, which makes sense as less data points could be used.

Based on these results, six models were selected to be included in the prediction pipeline: \textit{GAM}, \textit{ARIMA} and \textit{SARIMAX}, \textit{XGBoost}, \textit{CNN (Dilated)}, and \textit{HWES}.
Furthermore, we included \textit{Ensemble Mean}. 
We decided to include \textit{CNN (Dilated)} as the only deep learning approach in the pipeline, as it had the best prediction performance and computing multiple deep learning approaches would lead to higher computational costs. 
We excluded \textit{Facebook Prophet} for multiple reasons:
The model model delivered promising results, however, it could not outperform the other deep learning approaches. Further, it requires \texttt{rstan}, which was challenging to provide in the productive environment.
We included \textit{SARIMAX} as well as \textit{ARIMA} in the analysis pipeline as some potential future users of the pipeline might already use these models. The same holds for \textit{HWES}, which additionally had on average a bad performance due to strong bias in some predictions. 
We included \textit{XGBoost} and \textit{GAM} in the pipeline as the results were promising and the computational costs were comparably low.

\begin{table}[!ht]
    \centering
        \begin{tabular}{lrr}
        \toprule
        Method           &  \acs{nRMSE} &  \acs{MAPE}         \\
        \midrule
CNN (Dilated)             &  0.3119 &   33.19 \% \\
Tree-Based (XGBoost, log) &  0.3456 &   34.99 \% \\
Ensemble (median)         &  0.3693 &   42.91 \% \\
ARIMA (Seasonal)          &  0.3832 &   44.67 \% \\
ARIMA                     &  0.4105 &    53.0 \% \\
GAM                       &  0.4202 &   52.81 \% \\
Smoothing (HWES)          &  4.8962 &  236.79 \% \\
        \bottomrule
        \end{tabular}
        
    \caption{Selected models for analysis pipeline with \acf{nRMSE} and \acf{MAPE} averaged over all products.}
    \label{tab:show-case-error-models-selected}
\end{table}

Based on the results, we preselected seven models for the forecasting pipeline. 
Table \ref{tab:show-case-error-models-selected} shows the nRMSE and MAPE of the preselected models.
Notably, HWES has by far the worst performance averaged over all products.
This is due to outliers in the predictions, where HWES highly overestimate the the sales.
However, the average nRMSE as well as the MAPE were biased through the outlier and HWES could deliver reasonable results for multiple products.

\section{Automated Forecasting Pipeline}\label{sec:forecasting-pipeline}

\begin{figure}[!ht]
    \centering
    \includegraphics[width=\textwidth]{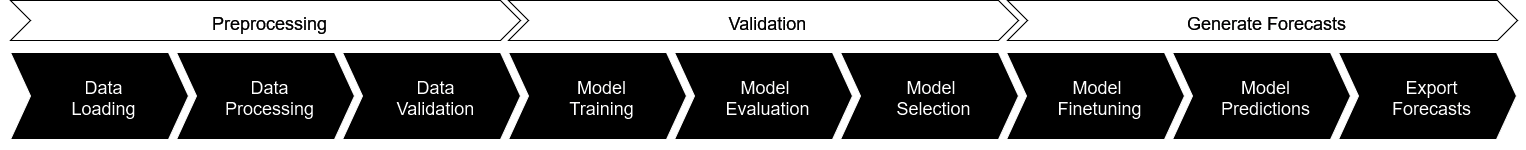}
    \caption{Illustration of the different steps of the analysis pipeline.}
    \label{fig:pipeline}
\end{figure}

In this section, we will start with an overview over the all the steps in the analysis pipeline including the automated model recommendation step. 
Afterwards, we will describe the methods used in the analysis. 
The analysis pipeline includes different sub tasks. 
After the data was loaded, it was checked for validity and processed in a specific format for weekly and monthly sales data as visualized in Figure \ref{fig:pipeline}.

In the pipeline, 6 models were trained and used for forecasting namely \ac{GAM}, \ac{ARIMA}, \ac{RF}, Deep Learning, and \ac{HWES}. 

\subsection{Step 1 - Preprocessing: Data Loading, Data Preprocessing and Data Checking}

First, the data was loaded from the \ac{ERP} system. 
The data contained previous sales aggregated by weeks and months for each product as separate time series.

Before starting with data processing, each time series was checked if enough data points were available. 
If a given time series contained less data then one year, e.g. for newly launched products, they were excluded in the automatic analysis. 

After valid time series were identified, data for the weekly and monthly predictions was processed. 
Instead of using hierarchical methods and predicting on a daily basis and aggregating the data afterwards, weeks and months were predicted independently to improve the performance. 
As the data is highly standardized through the \ac{ERP} system, data cleaning steps are not necessary except for format checks.
Additionally, no strategies were needed for missing values as the data was automatically collected.

\subsection{Step 2 - Validation: Model Training Model evaluation and Model Selection}

After preprocessing historical data, we trained various models using these data. 
We differentiated between models with shared weights and models individualized for each product. 
For the analysis, we employed deep dearning and random forest with shared weights for all products in our analysis. 
In contrast, auto regressive methods, such as \ac{ARIMA} and \ac{HWES}, were trained on a per-product basis.
At the first step, all models were trained on the historical data up to 1 year in the past and predicted the the remaining, untouched year. 
This validation delivered insights about the model performances. 
Based on that, the model with the lowest \ac{RMSE} will be recommended, which potentially will have the best prediction power for forecasting the future. 
As a error metrics, we used the root mean squared error defined as defined in Equation \ref{eq:RMSE}.

\subsection{Step 3 - Generate Forecasts: Model Finetuning, Prediction and Export}

After the model validation, all models were retrained including the test data.
These final models were used to predict the next 18 months or 78 weeks respectively. 
All models used in validation were used for prediction to allow the user selecting a different model manually.
The export included besides the forecast a summary over the predictions and the validation step to allow the user to get further information of the analysis results. 
Furthermore, plots were generated for each product visualising the decomposed time series in seasonality and trend to enable understanding of the data by the costumer.

\section{User Testing of the Final Forecasting Pipeline}\label{sec:user-test}

\subsection{Overview of the User group}

To evaluate the forecasting pipeline under real-world condition, we selected 5 different companies for testing based on the availability of their data and their expressed interest in utilizing the forecasting pipeline.
The companies A, B and C were specialized in the food and nutrition industry while company A had strong sales in winter compared to B and C. 
The companies  D and E were focusing on the electronic industry. Company D had the characteristic, that the sales were based on projects.

Each company was granted access to the pipeline and the users were getting a training on how to use the pipeline in production.
Within a test phase of half a year, we assumed that these companies would implement the pipeline as intended, providing meaningful insights into its effectiveness. 
For that, we have collected data of the predictions and the actual data to evaluate the automated pipeline. 
The baseline model for comparison was a naive model, which predicted future values based on the mean of previous years.
This model was commonly employed by the companies for planning purposes, making it a relevant and practical benchmark for our analysis.

\begin{table}[!ht]
    \centering
            
        \begin{tabular}{lcccll}
        \toprule
        Company &  Total Items &  Active Items & Active Items (\%) & First Date &  Last Date \\
        \midrule
        Company A &             2445 &           482 &          19.71 \% & 2014-01-01 & 2023-12-01 \\
        Company B &             4356 &          1234 &          28.33 \% & 2000-01-01 & 2024-01-01 \\
        Company C &              112 &           112 &          100.0 \% & 2014-04-01 & 2023-12-01 \\
        Company D &              783 &           345 &          44.06 \% & 2018-12-01 & 2023-12-01 \\
        Company E &             6317 &          3774 &          59.74 \% & 2006-01-01 & 2024-02-01 \\
        \bottomrule
        \end{tabular}
        
    \caption{Summary of the number of total items in the data set, the number of active items sold at least once in 2023, and the first and last data point.}
    \label{tab:tableOneOverviewUserGroup}
\end{table}

Table \ref{tab:tableOneOverviewUserGroup} provides a summary of the data obtained from the user group, highlighting variations among companies in terms of the number of items and the onset of data collection. 
For instance, Company D began collecting data in 2018 while Company E had the first data entry in 2006.
Furthermore, it is important to note that not all items sold at that time were projected to have sales figures in the subsequent year. 
Consequently, the dataset included items that have recorded zero sales in 2023. 
The number of \textit{Active Items} in Table \ref{tab:tableOneOverviewUserGroup} gives the count of items with sales in the year 2023, which were all used for validation.

\subsection{Experimental Set Up}

To assess the forecasting pipeline, we simulated its use starting from January 1\textsuperscript{st} 2023.
In this setup, the internal validation was performed using data from 2022, allowing us to compare the pipeline’s predictions with the actual sales data from 2023.
To quantify the model performance, we used the \ac{nRMSE} defined in Equation \ref{eq:RMSE}. This normalized error metric can be used for comparing predictions at different scales.

In our evaluation, we explored several key questions. 
Initially, we assessed the performance of the model recommended by our internal validation process by comparing the forecasts of the recommended model to the actual sales of the year 2023. 
Subsequently, we compared all models to determine which achieve highest predictive accuracy. 

We investigated how often a model could achieve the lowest \ac{nRMSE} in 2023. 
By comparing the selected models to the internal validation step with the model recommendation, we could evaluate how often we have selected the best model based on the internal validation data.
To evaluate the effectiveness of the model selection process critically, the results were compared to the naive model.
For visualizing the results, a box plot of the ratio was used.
Afterwards a formal test was used to compare the performances.
We used the Wilcoxon signed rank test by \cite{rey_wilcoxon-signed-rank_2011} to investigate if the recommended and best model performances are significantly better than the naive estimator. 

\subsection{Results}

\begin{figure}[!ht]
    \centering
    \includegraphics[width=0.7\textwidth]{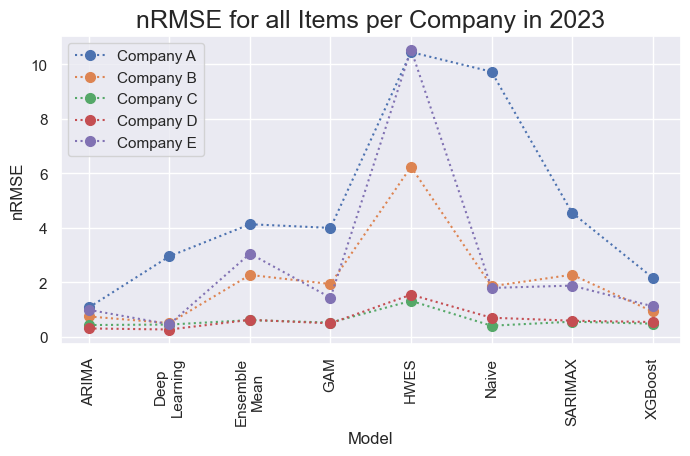}
    \caption{\acl{nRMSE} averaged over all products per model and company.}
    \label{fig:overview-nrmse}
\end{figure}

Figure \ref{fig:overview-nrmse} shows the average \ac{nRMSE} across all products for a company, to compare the model performances within and across the different companies.
The results showed that the model performance highly fluctuates among the companies, e.g. \ac{HWES} leads to a \ac{nRMSE} of 1.31 in Company C while in Company E the value of 10.54 is the highest. 
On average, Deep Learning had the lowest \ac{nRMSE} in Companies B, D, and E, while the lowest average error in company A and C was achieved with \ac{ARIMA}. 

\begin{figure}[!ht]
    \centering
    \includegraphics[width=\textwidth]{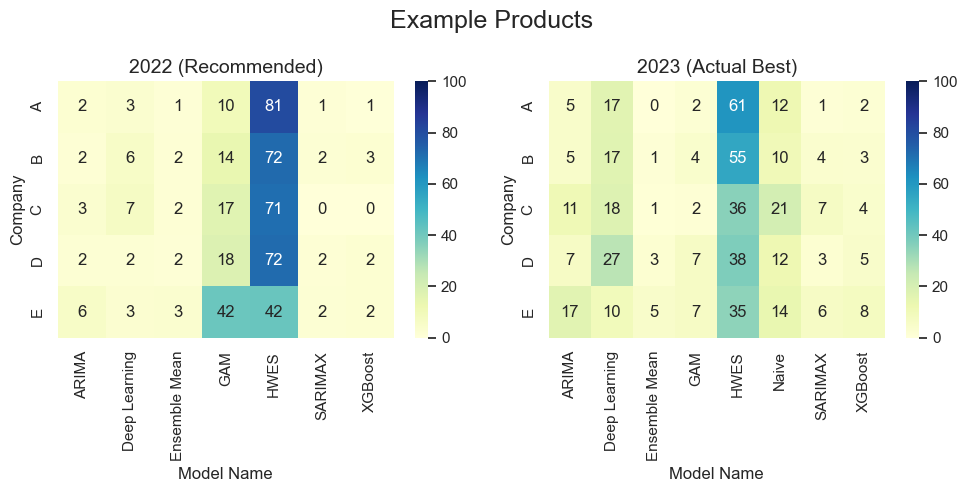}
    \caption{The plot shows a heatmap of how often the model was recommended based on the data from 2022 (left panel) and how often the model was actually the best in 2023 (right panel).}
    \label{fig:RecBestBarplot}
\end{figure}

The results in Figure \ref{fig:RecBestBarplot} show that based on the validation step, \ac{HWES} was recommended dominantly followed by the \ac{GAM}. Only in Company E, \ac{GAM} and \ac{HWES} were recommended equally often in 42\% of the products for the forecast. 
Surprisingly, the evaluation of the year 2023 showed that the actual best model with the lowest \ac{MAPE} differs from the recommended models. 
\ac{HWES} still seems to be the dominant model, but Deep Learning gained importance and performed in 35\% of the products for Company D (Company A 27\%, Company B 24\%) better.
Furthermore, we observed that the naive estimator was selected in 21\% of the cases for Company C.

\subsection{Relevance in Practice}

To elaborate the benefits of the proposed model pipeline, we compared our results of the analysis pipeline to a naive estimator using the mean of the previous data points as a prediction. 
For evaluation, we calculated the error ratio $r$ as the ratio of the \ac{nRMSE} of our model to the one of the Naive estimator: 

\begin{equation*}
    r = \frac{nRMSE_{Model}}{nRMSE_{Naive}}
\end{equation*}

\begin{figure}[!ht]
    \centering
    \includegraphics[width=\textwidth]{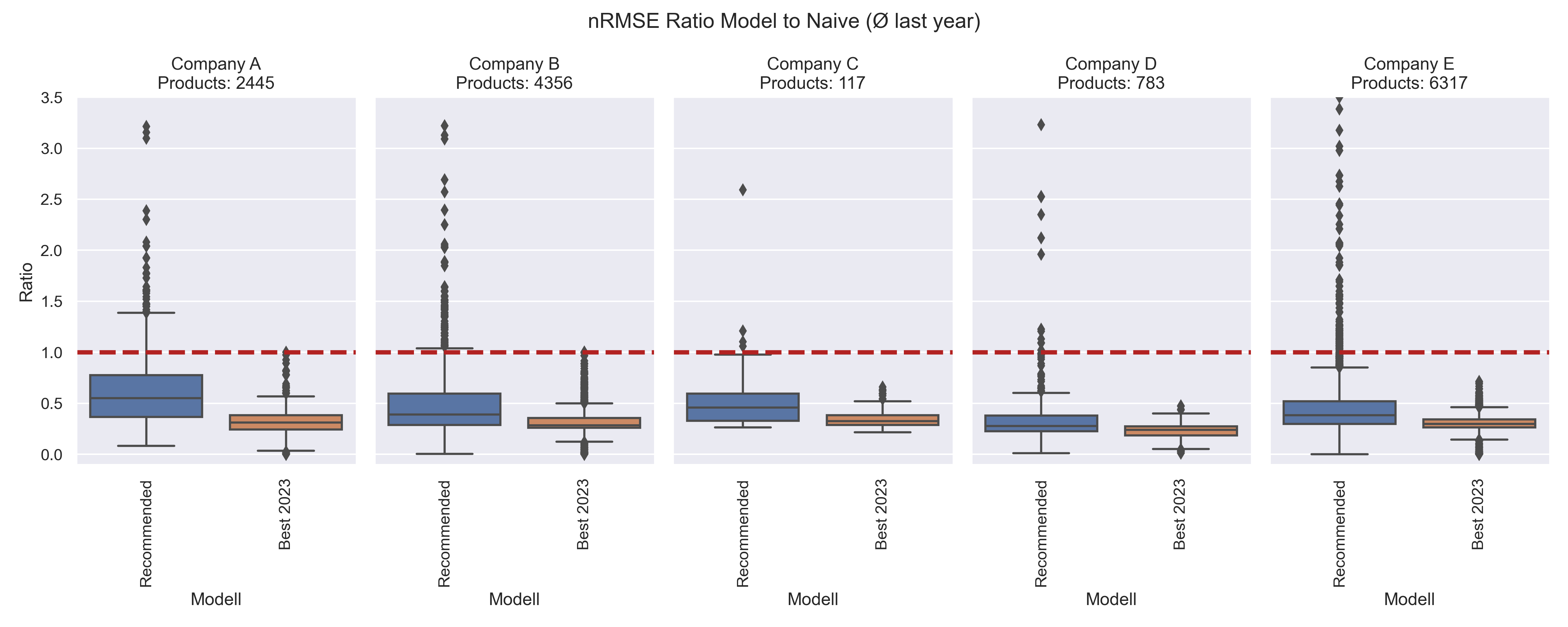}
    \caption{Error Ratio to naive estimator of the recommended model in the year 2023 and the actual best model. Note: The box plot was cut ad 3.5 while some outliers are greater then 3.5.}
    \label{fig:error-ratio}
\end{figure}

Here, $r < 1$ if the selected model outperforms the naive estimator.
Figure \ref{fig:error-ratio} presents boxplots visualizing the distribution of the error ratio between the recommended models and the naive estimator in different companies. 
Examining the results for the recommended model based on the validation of the previous year, we see the following results.



The boxplot for the best-performing model indicated that the recommended as well as the best models outperformed the naive estimator across all companies.
However, some recommended models performed worse compared to the naive estimator across all companies.
The quantiles on the boxplots provide an understanding of the variability and reliability of these results.

The statistical tests indicated that the differences in performance between the recommended and the naive estimator for companies A, C, D, and E were significant, while for Company B, the results were not statistically significant as shown in Table \ref{tab:pvalues}.
This confirms the visual interpretation of the box plot.

\begin{table}[!ht]
    \centering
        \begin{tabular}{lrr}
        \toprule
        {} & {Recommended Model} & {Best Model} \\
        {} &           p-Value &    p-Value  \\
        \midrule
            Company A &    1.41$ \times 10^{-6}$ &   1.05$ \times 10^{-95}$ \\
            Company B &              0.42 &  7.37$ \times 10^{-211}$ \\
            Company C &    9.64$ \times 10^{-8}$ &    1.63$ \times 10^{-9}$ \\
            Company D &   2.03$ \times 10^{-13}$ &   1.74$ \times 10 ^{-75}$ \\
            Company E &   9.16$ \times 10^{-27}$ &       0.00$ \times 10^{0}$ \\
        \bottomrule
        \end{tabular}

    \caption{Exact p-values for the Wilcoxon signed rank test indicating the difference of the recommended or best model compared to the naive estimator.}
    \label{tab:pvalues}
\end{table}

Overall, while the best models generally showed improved performance, they did not universally exceed the naive estimator's performance across all products, particularly in companies A, B, and C.
These findings highlight the need for further refinement of models for these companies. For companies D and E, the best models demonstrated the potential for consistent and significant improvement over the naive estimator, suggesting effective model application in these contexts.

\section{Discussion}\label{discussion}

This work presents an automated data analysis pipeline used in practice.
After identifying and presenting relevant literature, most promising models were preselected and implemented in an automated forecasting pipeline.
Next, the automated pipeline was tested in real world by five companies.
The results from the user test showed that the suggested data pipeline can outperform naive estimators by automatically recommending the right model in a validation step in 4 out of the 5 companies.

In our model pipeline, \acs{HWES} outperforms other models, which was not reflected in other research, where deep learning approaches or autoregressive models delivered the best performances. \citep{hamzacebi_comparison_2009, hassan_el_madany_procurement_2022}

However, since the validation step did not identify the best model in all products, further research has to be done. 

As the data was collected within an \ac{ERP} system and was exported with all available features, the data was highly standardized and did not need individual preprocessing or feature extraction. With that, this approach might be easily scaleable and adoptable.
Nevertheless, the history itself might be corrupt.
For instance, we saw that data maintenance within the company could issue sales which cannot be allocated to the product identifiers, leading to lower sales in the history. 
That happens, if items changed their identifier without a data entry or note in the \ac{ERP} system. 
Furthermore, the data collected in the \ac{ERP} system were sales figures. 
Data about rare out-of-stock events or delayed deliveries are not collected, but might influence the practical relevance.
With that, the data may not match the actual demand.

Within this study, we used a prediction horizon of 1.5 years.
As biases increase with longer-term predictions, the forecasts must be handled with care.
Unforeseen events or trend change points can highly bias the predictions. 
This effects the validation step as well, as the environmental setting can change rapidly between the validation period and the prediction period. 
With that, the models recommended based on the validation period can be totally different from the actual best models. 

We saw, that in the first model selection approach, Deep Learning demonstrated strong performance. 
However, we were unable to reproduce these results in the pilot study. 
Several factors could contribute to this discrepancy.
Firstly, the model selection dataset included 51 manually pre-selected, well-established products, ensuring high data quality.  
In contrast, the pilot study utilized real-world data from the \ac{ERP} system, which is inherently lower in quality due to shorter histories, outliers, and unmatched identifiers.
Despite efforts to reduce bias and exclude extreme item histories in the data pipeline, achieving the same quality level as the pre-selected dataset proved challenging.
Additionally, statistical analysis indicates significant variability in model performance due to the aforementioned data quality issues. 
The robustness and generalizability of the models are also impacted, highlighting the limitations of selecting models on the previous year in real-world datasets.

In this paper, we were focusing on the historical data. Further research can include additional features like marketing expanses or general information like market growth, which might have an huge influence on the model performances. 
Furthermore, future work could focus on improving data preprocessing and validation techniques and exploring a robust model selection approach that are less sensitive to data quality variations.

In conclusion, this research provides a thorough assessment of an automated demand forecasting pipeline, demonstrating its potential to address key real-world challenges. Additionally, it offers valuable insights into model performance and introduces a viable approach for automated model evaluation, paving the way for more efficient and scalable forecasting solutions.


\begin{acronym}
 \acro{ARMA}{autoregressive moving average}
 \acro{ARIMA}{autoregressive integrated moving average}
 \acro{CNN}{Convolutional Neural Network}
 \acro{ERP}{Enterprise-Resource-Planning}
 \acro{ES}{Exponential Smoothing}
 \acro{GAM}{Generalized Additive Model}
 \acro{GRU}{Gated Recurrent Unit}
 \acro{HWES}{Holt-Winters Exponential Smoothing}
 \acro{LSTM}{long short-term memory}
 \acro{MAPE}{Mean Absolute Percentage Error}
 \acro{MLP}{Multilayer Perceptron}
 \acro{RF}{Random Forest Regressor}
 \acro{RMSE}{Root-Mean-Squared Error}
 \acro{nRMSE}{normalized Root-Mean-Squared Error}
 \acro{RNN}{Recurrent Neural Network}
 \acro{SME}{small- to medium-sized enterprises}
 \acro{STD}{Standard Deviation}
 \acro{XGBoost}{Extreme Gradient Boosting}
\end{acronym}

\section*{Acknowledgements}


We thank all the cooperation partners that supported the publication with their data insights. Furthermore, we thank Thomas Kluth, Antje Heine, Hans-Peter Hantsch,  Katharina Kluth from abacus edv-lösungen for their experience from industry, technical support, and Juliana Schneider from Hasso-Plattner-Institute for scientific discussions.

\section*{Funding sources}

This work received funding from \textit{abacus edv-lösungen GmbH \& Co. KG}.

\section*{Competing Interests}

The authors declare that this work received funding from \textit{abacus edv-lösungen GmbH \& Co. KG}, which also uses the described pipeline commercially. However, the research and findings were conducted independently and were not influenced by financial or commercial interests.

\bibliographystyle{unsrt}  

\bibliography{Zotero}

\newpage
\renewcommand{\tablename}{Supplementary Table}
\section*{Appendix}

\begin{table}[!ht]
    \centering
    \begin{small}
        \begin{tabular}{l|ll|ll}
        \toprule
                   &  \multicolumn{2}{c}{\acs{nRMSE}} &  \multicolumn{2}{|c}{\acs{MAPE}}         \\
        Method & Mean & \acs{STD} & Mean & \acs{STD} \\
        \midrule
\textit{CNN (Dilated)  }                               &        0.3119 &       0.1132 &     33.19 \% &      0.1623 \\
CNN                                           &        0.3121 &       0.1403 &     37.01 \% &      0.2399 \\
CNN (pretrained)                              &        0.3138 &       0.1468 &     35.07 \% &      0.2123 \\
CNN (FIXED)                                   &        0.3247 &       0.1621 &     36.83 \% &      0.2515 \\
\textit{Tree-Based (XGBoost, log)}                     &        0.3456 &       0.1571 &     34.99 \% &      0.1481 \\
Tree-Based (XGBoost)                          &        0.3463 &       0.1876 &     36.64 \% &      0.1858 \\
GRU                                           &        0.3498 &       0.1897 &     47.33 \% &      0.4430 \\
LSTM (3)                                      &        0.3573 &       0.1082 &     45.01 \% &      0.1759 \\
CNN (DAY)                                     &        0.3574 &       0.1312 &     36.96 \% &      0.1674 \\
MLP                                           &        0.3642 &       0.2292 &     42.46 \% &      0.3151 \\ \hline
LSTM (1)                                      &        0.3649 &       0.0622 &     40.05 \% &      0.1371 \\
\textit{Ensemble (median)}                             &        0.3693 &       0.2038 &     42.91 \% &      0.2964 \\
Ensemble (median, dependent)                  &        0.3757 &       0.2055 &     44.27 \% &      0.3071 \\
\textit{ARIMA (Seasonal)}                              &        0.3832 &       0.2191 &     44.67 \% &      0.3232 \\
Facebook-Prophet (Day)                        &        0.3873 &       0.3495 &      50.2 \% &      0.5405 \\
ARIMA (Seasonal, median, dependent)           &        0.3906 &       0.2158 &     46.18 \% &      0.3207 \\
ARIMA (Seasonal, mean, dependent)             &        0.3924 &       0.2151 &     45.89 \% &      0.3229 \\
ARIMA (median, dependent)                     &        0.3982 &       0.1947 &     48.78 \% &      0.2897 \\
ARIMA (mean, dependent)                       &        0.3987 &       0.1947 &     49.28 \% &      0.2872 \\
CNN (FIXED, DAY)                              &        0.4026 &       0.4156 &     41.35 \% &      0.3529 \\
Tree-Based (Random-Forest, median, dependent) &        0.4028 &       0.1952 &     47.92 \% &      0.2973 \\
Tree-Based (Random-Forest)                    &        0.4029 &       0.1910 &     48.91 \% &      0.2959 \\
Tree-Based (Random-Forest, mean, dependent)   &        0.4042 &       0.1955 &     47.74 \% &      0.2891 \\
\textit{ARIMA}                                         &        0.4105 &       0.1860 &      53.0 \% &      0.2990 \\
\textit{GAM}                                           &        0.4202 &       0.2577 &     52.81 \% &      0.3089 \\
LSTM (2)                                      &        0.5097 &       0.2448 &     60.09 \% &      0.3311 \\
Facebook-Prophet                              &        0.5164 &       0.6405 &     76.06 \% &      1.1999 \\
Smoothing (HWES, median, dependent)           &        0.5582 &       0.7761 &     64.59 \% &      0.8684 \\
Ensemble (mean, dependent)                    &        0.6119 &       1.5512 &     53.27 \% &      0.5645 \\
Smoothing (ES)                                &        0.6498 &       0.5617 &      93.4 \% &      0.5976 \\
PytorchTS                                     &        0.7556 &       0.3324 &     86.08 \% &      0.0644 \\
Smoothing (HWES, mean, dependent)             &        1.3867 &       6.2870 &     85.22 \% &      1.9953 \\
Ensemble (mean)                               &        1.4689 &       6.4076 &     89.94 \% &      1.9253 \\
\textit{Smoothing (HWES)}                              &        4.8962 &      25.8723 &    236.79 \% &      7.7315 \\
        \bottomrule
        \end{tabular}
    \end{small}
    \caption{All Model Performances sorted by the mean of the \acf{nRMSE}. Furthermore, the \acf{STD} and \acf{MAPE} are presented.
    After the top-10, there is a horizontal line and models used for the prototype are highlighted italic.}
    \label{sup-table:error-all}
\end{table}

\end{document}